%Paper: cond-mat/9210001
%From: KETOJA@phcu.helsinki.fi
%Date: Thu, 1 Oct 1992 09:03 +0300

\magnification=1200
\pretolerance=1000
\parindent 5pt
\parskip=0.7cm
\hsize 12.5cm
\baselineskip=0.7cm

\line{\hfil}
\vskip 2cm
\line{\hfil {\bf UNIVERSAL CRITERION FOR THE BREAKUP OF INVARIANT}\hfil}
\line{\hfil {\bf TORI IN DISSIPATIVE SYSTEMS}\hfil}
\vskip 1cm
\line{\hfil Jukka A. Ketoja\hfil}
\medskip
\line{\hfil Research Institute for Theoretical Physics,\hfil}
\line{\hfil University of Helsinki,\hfil}
\line{\hfil Siltavuorenpenger 20 C,\hfil}
\line{\hfil SF-00170 Helsinki, Finland\hfil}
\bigskip
\line{PACS number: 05.45.+b \hfil}
\vskip 1cm
{\centerline{\bf Abstract}}

The transition from quasiperiodicity to chaos is studied in
a two-dimensional dissipative map with the inverse golden mean
rotation number. On the basis of a decimation scheme,
it is argued that the (minimal) slope of the critical iterated
circle map is proportional to the effective Jacobian determinant.
Approaching the zero-Jacobian-determinant limit, the factor of proportion
becomes a universal constant.
Numerical investigation on the dissipative standard map
suggests that this universal number could become observable
in experiments.

\vfill\eject

In an almost linear continuous dynamical system with two competing frequences,
the attractor is typically
either a mode-locked periodic state or an invariant torus associated
with quasiperiodic motion. In the parameter space, periodic
and quasiperiodic attractors are mingled up in such a way
that there is a positive probability for each kind of orbit
to be found. Experiments indicate that mode-locking
and quasiperiodic behaviour are generic in
hydrodynamics [1], charge-density-wave conductors [2] and
other physical [3,4] and chemical [5] systems.
Changing parameter values may increase
the nonlinearity and lead to a transition to chaos.

The experiment by Martin and Martienssen [3] on the electrical
conductivity of barium sodium niobate crystals is a very nice
example of the case in which it is possible to measure the
actual return map characterizing the discretized dynamics on an
invariant circle (the invariant torus appears as an invariant
circle for the Poincar\'e map of the system).
Bohr et al. [6,7] point out the intimate connection between
the existence of an invariant circle and the
one-dimensional nature of the return map. In particular, they show that
a zero slope in the return map is impossible if the underlying invariant
circle is smooth. The fact that the invariant circle
loses smoothness before breaking up [8,9]
could mislead one into thinking that a zero slope in the return map
is a necessary condition for the
system to be critical, i.e. about to become chaotic. Another motivation for
this kind of false idea could come from the fact that
an analytic circle map
has a zero-slope inflection point at the transition from
quasiperiodicity to chaos [10]. However, if the Jacobian determinant of the
Poincar\'e map is positive everywhere along the invariant circle,
it is impossible that the "reduced" circle map, i.e. the
projection of the Poincar\'e map on the invariant
circle, would have a zero slope at some point of the circle.
The reduced circle map could
have a zero slope only if the tangent vector
at the corresponding point was annihilated
by the Jacobian matrix. This could happen only if the Jacobian
determinant vanished. The positivity of the slope of
a critical circle map has been noticed by several
authors [11].

In this letter the relation between the slope of the reduced circle map
and the Jacobian determinant is elaborated further.
For simplicity, I will restrict myself to a
two-dimensional Poincar\'e map with rotation number $\zeta =(\sqrt 5 -1)/2$.
The reduced circle map is
denoted by $h(x)$ where $x$ is a scaled "angle" variable for
the invariant circle so that $h(x+1)=h(x)+1$. By the assumption
on the rotation number, $h^n (x)/n \to \zeta$ and, moreover,
$h_n (x) -F_{n-1} \equiv$$h^{F_n} (x) -F_{n-1}$$\to x$ as $n$ tends to
infinity.
Here $F_n$ stands for the $n$th Fibonacci number,
$F_{n+1} =F_n + F_{n-1}$ ($F_0 =0,\;F_1 =1$).
The Jacobian determinant of the $F_n$ times iterated
map at $x$ is denoted by $J_n (x)$.
It will be shown below that in the critical case $h_n '(x_0 ) \sim J_n (x_0 )$,
where $x_0 $ is a special point [8,9] associated with the universal scaling
by $\alpha \approx -1.2885746$. $x_0$ corresponds to a cubic
critical point for an analytic circle map. In a higher dimensional dissipative
system, $x_0$ can be searched either as the point visited most rarely by
the quasiperiodic orbit or as the limit of points
$x_n$, $n\to \infty$,  such that $h_n '(x)$ has a minimum at $x_n$.
It could as well be stated that $h_n '(x_n ) \sim J_n (x_n )$.
For a dissipative system, $J_n (x)\to 0$ as $n\to \infty$
so that a zero slope is indeed observed but only considering
the limit of an infinitely high iterate of the original circle map.

Furthermore, the calculation shows that
the factor of proportion between $h_n '(x_0 )$ and $J_n (x_0 )$
tends to a universal constant as the Jacobian determinant
approaches zero. In this limit,
$${h_n ' (x_0 ) \over J_n (x_0 )} \to a=
{2\over \lbrack \eta (0) \rbrack^2 \eta '''(0)} \hbox{ as }n\to \infty
\eqno(1)$$
where $\eta (x)$ is one of the components of the universal
fixed point pair $(\xi ,\eta )$ for the standard renormalisation operator
$T(\xi ,\eta )=\alpha (\eta ,\eta \circ \xi )\alpha^{-1}$ for
analytic circle maps [8].
MacKay's [12] expansion for $\eta (x)$
leads to the numerical estimate $a\approx 0.435625$.

The starting point is Bohr's [7] formula relating the derivatives of
the first and the second iterate
of the reduced circle map. Consider
a two-dimensional map $G(x,y)=(g_1 (x,y), g_2 (x,y))$
with an invariant circle $y=c(x)$ ($G(x+1,y)=G(x,y)+(1,0)$).
The original map can be related to
a one-dimensional circle map by $h(x)=g_1 (x,c(x))$ and
$c(h(x))=g_2 (x,c(x))$. Differentiating these two equations
with respect to $x$ yields, after some manipulation,
$${dh^2 (x)\over dx}=\lbrack g_{11} (h(x)) +{g_{22} (x)g_{12} (h(x)) \over
g_{12} (x)}\rbrack h'(x) -{g_{12} (h(x))\over g_{12} (x)} J(x) \eqno(2)$$
where
$$\eqalign{ g_{i1} (x)&={\partial g_i (x,y) \over \partial x}_{|y=c(x)}
,\;\;\; g_{i2} (x)={\partial g_i (x,y) \over \partial y}_{|y=c(x)} \cr
J(x)&= g_{11} (x) g_{22} (x) -g_{12} (x) g_{21} (x) \cr} $$
It is important that the
equation for the invariant circle does not appear in (2).

Eq. (2) can be written in the form
$$h_3 '(x)=c_2 (x)h_2 '(x) + d_2 (x) \eqno(3)$$
I introduce a decimation technique
by which one can generate from (3) and the trivial equation
$h_2 '(x)= 1h_1 '(x)+0$ a sequence of equations
$$h_{n+1} '(x)=c_n (x)h_n '(x)+ d_n (x) \eqno(4)$$
with increasing $n$.
Assume that (4) is known for $n=i$ and $n=i-1$.
Write $h_{i+2} '(x)$ as
$$h_{i+1} '(h_i (x))h_i '(x)
=c_i (h_i (x))h_i '(h_i (x))h_i '(x)+d_i (h_i (x))h_i '(x)\eqno(5)$$
and split $h_i '(h_i (x))$ further:
$$h_i '(h_i (x)) =c_{i-1} (h_i (x))h_{i-1} '(h_i (x))
+d_{i-1} (h_i (x)) \eqno(6)$$
Replacing $h_{i-1} '(h_i (x))h_i '(x)$ by $h_{i+1} '(x)$ and using
the fact that
$$h_i '(x)={h_{i+1} '(x)-d_i (x)\over c_i (x)} $$
leads finally to
$$h_{i+2} '(x)=c_{i+1} (x)h_{i+1} '(x)+d_{i+1} (x)$$
with
$$\eqalign{c_{i+1} (x)&=c_i (h_i (x))c_{i-1} (h_i (x))
-{d_{i+1} (x)\over d_i (x)} \cr
d_{i+1} (x)&=-{d_i (x)\lbrack d_i (h_i (x))+c_i (h_i (x))d_{i-1} (h_i (x))
\rbrack \over c_i (x)} \cr} \eqno(7)$$

These recursion relations help in determining the leading asymptotic
behaviour of $c_n (x)$ and $d_n (x)$ as $n$ tends to infinity.
First, it can be inductively argued that $d_n (x) \sim J_n (x)$
(except for $n=1$). Recall that $d_1 (x)\equiv 0$ and $d_2 (x)\sim J(x)$
so that (7) implies $d_3 (x)\sim J(x)J(h(x))=J_3 (x)$.
At each level $n\geq 3$ of the recursion, the leading term in the Jacobian
determinant arises from the product $\sim d_n (x)d_{n-1} (h_n (x))$
$\sim J_n (x)J_{n-1} (h_n (x))$$=J_{n+1} (x)$.
One can now proceed to determine the asymptotic
behaviour of $c_n (x)$ as $n\to \infty$. Eqs. (5-7) give
$$h_{n+1} '(h_n (x))=c_{n+1} (x) h_{n-1} '(h_n (x))
+{d_{n+1} (x)\over d_n (x)} \lbrack h_{n-1} '(h_n (x))-c_n (x)\rbrack $$
Because $d_{n+1} (x)/d_n (x) \to 0$,
all one needs to know is the asymptotic behaviour
of $h_{n+1} '(h_n (x))$ and $h_{n-1} '(h_n (x))$.
I consider here only
the critical case with $x=x_0 $.
On the basis of the renormalisation theory [8,9,13], it is expected that
$$h_n (x_0 +z)-F_{n-1} -x_0 \approx \alpha^{-n} \eta( \alpha^n z) \eqno(8)$$
where the approximation improves with increasing $n$.
This implies
$h_{n+1} '(h_n (x_0 ))$\hfill\break $\to \eta '(\xi (0))=\alpha^4$ and
$h_{n-1} '(h_n (x_0 )) \to \xi '(\eta (0))=\alpha^2$ where
the derivatives have been calculated from the fixed point
equation. Thus, $c_n (x_0 )\to \alpha^2 \approx 1.66$
as $n$ tends to infinity.

The asymptotic
behaviour of $h_n '(x_0)$
is solely determined by those of $c_n (x_0 )$ and $d_n (x_0 )$.
If $h_n '(x_0 )$ approached zero slower than
$d_n (x_0 )$, there would exist an $N$ such that for all $n>N$,
$h_{n+1} '(x_0 )>C h_n '(x_0 )$ with $C>1$. In this case,
$h_n '(x_0 )$ would actually keep growing without any limit
as $n \to \infty$ which would be
contradictory to the tendency of the renormalised circle map to approach
the universal function $\eta (x)$.
On the other hand, (4) implies that $h_n '(x_0 )$ cannot
decay to zero faster than $d_n (x_0 )$. In other words,
$$h_n '(x_0 )=-{d_n (x_0)\over c_n (x_0 )}+O(J_{n+1} (x_0 ))
\sim J_n (x_0 ) $$

It turns out to be possible to work out the limit of the factor
$$e_n (x)= -{d_n (x) \over c_n (x) J_n (x)} $$
at $x=x_0$ approaching the case in which the Jacobian determinant
vanishes. Note first that
$$e_2 (x)={g_{12} (h(x)) \over g_{12} (x) c_2 (x)},
\;\;e_3 (x)={g_{12} (h_3 (x)) \over g_{12} (x) c_2 (x) c_3 (x)} $$
Eq. (7) implies a recursion relation for $e_n (x)$, $n=3,4,...$
which becomes very simple in the zero-Jacobian-determinant limit:
$$e_{n+1} (x) \approx e_n (x) e_{n-1} (h_n (x)) $$
Because $c_3 (x)$ can be replaced by
$c_2 (h_2 (x)) c_1 (h_2 (x))\equiv c_2 (h_2 (x))$,
it is easy to write down the form of a general $e_n (x)$:
$$e_n (x)\approx
{g_{12} (h_n (x)) \over g_{12} (x) \prod_{i=0}^{F_n -1}
c_2 (h^i (x))} \approx {1 \over \prod_{i=0}^{F_n -1}
c_2 (h^i (x))} $$
Here I have used the fact that $h_n (x)$ mod $1$$\to x$
as $n\to \infty$ for
the inverse golden mean rotation number.
Leaving the $d$-term proportional to the
Jacobian out of (4), one obtains
$$\prod_{i=0}^{F_n -1} c_2 (h^i (x)) = h_n '(h(x)) $$
where the derivative can be calculated using (8):
$$h_n '(h(x_0 +z))
\approx  {h'(x_0 +\alpha^{-n} \eta (\alpha^n z))\eta '(\alpha^n z) \over
h'(x_0 +z)} $$
$\eta (z)$ has a cubic critical point at $z=0$ [8]. Furthermore, also
$h (x_0 +z)$ developes such a point in the zero-Jacobian-determinant
limit.
Expanding all the derivatives around $z=0$
and letting $z\to 0$ leads to
$$h_n '(h(x_0 ))\to {\lbrack \eta (0) \rbrack^2 \eta '''(0) \over 2} $$
We have thus derived Eq. (1).

Table 1 shows $h_n '(x_0 )/J_n $
for the dissipative standard map
$$\eqalign{g_1 (x,y)&=x+\Omega+by-{k\over 2\pi} \sin (2\pi x)\cr
g_2 (x,y)&=\Omega+by-{k\over 2\pi} \sin (2\pi x)\cr} $$
with the constant Jacobian determinant $b=0.5$. The critical parameter values
for the breakup of the "golden" invariant circle can be determined
by a dissipative version of Greene's residue criterion [14].
$x_0 ,y_0 $ is taken as the point where the approximating
periodic orbits have the largest gap.
The calculation of $h_n '(x_0 )/J_n $ using the forward recursion relation (4)
would be extremely sensitive to the choice of the value of
$h_2 '(x_0 )\equiv h'(x_0 )$. An error $\epsilon$ in
$h'(x_0 )$ would give
 rise to an error $\epsilon \prod_{i=2}^{n-1} c_i ( x_0 )$
in $h_n '(x_0 )$ which would be of the order $\epsilon \; 1.66^{n-2}$. As $J_n$
decays to zero very fast with increasing $n$ ($J_{15} \sim 10^{-184}$),
the error in the ratio $h_n '(x_0 )/J_n$ would soon become
astronomical. A better way to calculate this ratio is to apply (4)
backwards beginning with the approximation $h_N '(x_0 )\approx 0$ for
some large $N$. The initial error is very small ($\sim J_N $)
and the error in $h_n '(x_0 )$ is reduced
by a factor around $1.66$ at each step. In fact, a very good
estimate for the ratio $h_n '(x_0 )/J_n $ is obtained already for
$n=N-1$ if $N$ is not very small.
The calculation of $c_n (x_0 )$ and $d_n (x_0 )$ by
(7) appears to be numerically very stable. Thus, all the error
arises essentially from the inaccuracy in determining the critical
parameter values and the point $x_0 ,y_0$.

Table 1 shows no deviation from (1) although the system is
quite far from the zero-Jacobian-determinant limit. This could
be taken as a hint that Eq. (1) would be valid more generically
than the derivation would reveal.
It would be intriguing to see this tested experimentally.
If the experimental data enabled one to construct the one-dimensional
return map, it would be quite easy to calculate derivates
of higher iterates of this return map by using finite differences.
Usually the point $x_0$ and the Jacobian determinant
would not be known. It would be best to
estimate the smallest slope of each Fibonacci iterate of the reduced
circle map and
calculate $h_n ' h_{n-1} ' / h_{n+1} '$. If the Jacobian determinant
varied only little along the invariant circle, this ratio could
be close to the universal constant $a$. It is clear that noise
would prevent one from carrying out the calculation for high $n$.
Nevertheless, Table 1 suggests that even the lowest order
estimate could give a reasonable result.

 As to other rotation
numbers, I would expect
the effective Jacobian again to play an important role [15]. The transition
to chaos should be observed by
monitoring the smallest slope of the higher iterate $h^{Q_n} (x)$
 of the
reduced circle map, where $Q_n$ would be the denominator of the $n$th
truncation of the continued fraction expansion for the rotation number.
In the critical case, one would expect this slope to
tend to zero as $n\to \infty$
whereas in the subcritical region the asymptotic slope should be unity [8,9].

Eqs. (4) and (7) can be used
to study the conservative case $J(x)\equiv 1$ as well. Both
$c_n (x_0 )$ and $d_n (x_0 )$  have universal non-vanishing limits
($x_0$ now corresponds to a point
on a dominant symmetry line [12]):
$c_{\infty} (x_0 ) \approx 2.1676633$ and $d_{\infty} (x_0 )\approx
-0.4916138$.
$c_{\infty} (x_0 )$ appears to be determined by the ratio
of the universal phase space scaling constants [12].
$h_n '(x_0 )$ has a
universal positive limit $d_{\infty} (x_0 )/(1-c_{\infty} (x_0 ))
\approx 0.4210236$. The estimate obtained by setting $J_n =1$ in (1)
deviates only about
$3$ \% from this true value.

The decimation technique introduced in this
letter is readily applicable also to other problems, e.g. to the
discrete quasiperiodic Schr\"odinger equation [16].
Eq. (3) can be interpreted as a discrete
eigenvalue equation with $h_n '$ representing
the wave vector $\psi_{F_n}$ at site $F_n$ [17].
The modulating potential is included in $c_2$. Assuming
the normalisation condition $\psi_0 =1$ one can take $d_2 \equiv -1$.
The present approach
is appropriate when the potential has the frequency $\zeta$
relative to the underlying lattice. Infinite products of transfer
matrices usually diverge [18] whereas, by writing down recursion relations
similar to (7), it is possible to find a bounded limiting
behaviour for the coefficients $c_n$ and $d_n$ [17].

The fact that the slope of the reduced circle map
depends on the effective Jacobian is in a
nice agreement with the conjectured mechanism for the breakup of
having a tangency between the invariant circle and
its stable foliation [8]. It is natural to think the contraction on the stable
foliation to be proportional to the effective Jacobian determinant.
At the point of tangency, the slope of the associated circle
map should therefore be proportional to the Jacobian.

I would like to thank M.H. Jensen for pointing out Ref. [3] to me and
J. Kurkij\"arvi for comments on the manuscript.

\vfill\eject

{\bf References}

[1] J. Stavans, F. Heslot and A. Libchaber,
Phys. Rev. Lett. 55, 596 (1985); A.P. Fein, M.S. Heutmaker and
J.P. Gollub, Physica Scr. T9, 79 (1985);
D.J. Olinger and K.R. Sreenivasan, Phys. Rev. Lett. 60, 797 (1988);
R.E. Ecke, R. Mainieri and T.S. Sullivan,
Phys. Rev. A 44, 8103 (1991).

[2] S.E. Brown, G. Mozurkewich and G. Gr\"uner,
Phys. Rev. Lett. 52, 2277 (1984);
S. Bhattacharya, M.J. Higgins and J.P. Stokes,
Phys. Rev. B 38, 7177 (1988).

[3] S. Martin and W. Martienssen, Phys. Rev. Lett. 56, 1522 (1986).

[4] G.A. Held and C. Jeffries, Phys. Rev. Lett. 56, 1183 (1986);
E.G. Gwinn and R.M. Westervelt, {\it ibid.} 57, 1060 (1986);
A. Cumming and P.S. Linsay, {\it ibid.} 59, 1633 (1987);
M. Bauer, U. Krueger and W. Martienssen, Europhys. Lett. 9,
191 (1989); J. Peinke, J. Parisi, R.P. Huebener,
M. Duong-van and P. Keller, {\it ibid.} 12, 13 (1990);
W.J. Yeh, D.-R. He and Y.H. Kao, Phys. Rev. Lett. 52, 480 (1984);
P. Alstr{\o}m and M.T. Levinsen, Phys. Rev. B 31, 2753 (1985).

[5] J. Maselko and H.L. Swinney, J. Chem. Phys. 85, 6430 (1986).

[6] T. Bohr, P. Bak and
M.H. Jensen, Phys. Rev. A 30, 1970 (1984);
P. Bak, T. Bohr, M.H. Jensen and P.V. Christiansen,
Solid State Commun. 51, 231 (1984).

[7] T. Bohr, Phys. Lett. A 104, 441 (1984).

[8] S. Ostlund, D. Rand, J. Sethna and E. Siggia, Physica D 8,
303 (1983).

[9] M.J. Feigenbaum, L.P. Kadanoff and S.J. Shenker,
Physica D 5, 370 (1982).

[10] R.S. MacKay and C. Tresser, Physica D 19, 206 (1986).

[11] X. Wang, R. Mainieri and J.H. Lowenstein, Phys. Rev. A 40,
5382 (1989); S.-Y. Kim and B. Hu, {\it ibid.} 44, 934 (1991).

[12] R.S. MacKay, Ph.D. Thesis, University of Princeton (unpublished,
1982).

[13] D. Rand, in New directions in dynamical systems, eds.
T. Bedford and J. Swift (Cambridge University Press, Cambridge, 1988)
p. 1.

[14] J.A. Ketoja, Physica D 55, 45 (1992).

[15] The effective Jacobian controls the "universal crossover"
 in period-doubling systems as well; see
C. Reick, Phys. Rev. A 45, 777 (1992), and references therein.

[16] For a review, see J.B. Sokoloff, Phys. Rep. 126, 189 (1985).

[17] J.A. Ketoja, to be published.

[18]  S. Ostlund and R. Pandit, Phys. Rev. B 29, 1394 (1984).

\vfill\eject

Table 1. Subsequent estimates for the factor $h_n '(x_0 )/J_n$ resulting from
the critical dynamics of the dissipative standard map
($b=0.5$, $k=0.9788377790$, $\Omega=0.3058769514$).

$$\vbox{\offinterlineskip
\hrule
\halign{&\vrule #&\strut\quad \hfil# \quad\cr
&$n$&&$h_n '(x_0 )/J_n $&\cr
\noalign{\hrule}
&$2$&&$0.5280$&\cr
&$3$&&$0.4333$&\cr
&$4$&&$0.4548$&\cr
&$5$&&$0.4146$&\cr
&$6$&&$0.4384$&\cr
&$7$&&$0.4191$&\cr
&$8$&&$0.4385$&\cr
&$9$&&$0.4272$&\cr
&$10$&&$0.4387$&\cr
&$11$&&$0.4312$&\cr
&$12$&&$0.4380$&\cr
&$13$&&$0.4331$&\cr
&$14$&&$0.4373$&\cr
&$15$&&$0.4339$&\cr
&$ $&&$ $&\cr}
\hrule}$$

\end